\documentclass[%
 reprint,
superscriptaddress,
 amsmath,amssymb,
 aps,
prb,
]{revtex4-2}

\usepackage{graphicx}
\usepackage{dcolumn}
\usepackage{bm}
\usepackage{textcomp}

\usepackage{xcolor}

\usepackage{newtxtext}

\begin{document}

\preprint{APS/123-QED}

\title{Geometry of the vapor layer under a Leidenfrost hydrogel sphere}

\author{Vicente L. Diaz-Melian} 
\email{vicenteluis.diazmelian@ista.ac.at}
\affiliation{%
 Institute of Science and Technology Austria, Am Campus 1, 3400 Klosterneuburg, Austria
}%

\author{Isaac C. D. Lenton}

\affiliation{%
Institute of Science and Technology Austria, Am Campus 1, 3400 Klosterneuburg, Austria
}%
\author{Jack Binysh}%
\affiliation{
Institute of Physics, Univ. Amsterdam, Science Park 904, 1098 XH Amsterdam, The Netherlands
}%
\author{Anton Souslov}
\affiliation{
TCM Group, Cavendish Laboratory, JJ Thomson Avenue, Cambridge, CB3 0HE UK 
}%

\author{Scott R. Waitukaitis}
\affiliation{%
Institute of Science and Technology Austria, Am Campus 1, 3400 Klosterneuburg, Austria
}%


\date{\today}

\begin{abstract}

A floating Leidenfrost droplet exhibits curvature inversion of its underside, due to the balance of vapor pressure and surface tension. Using interferometric imaging, we find different behavior for a levitated hydrogel sphere. Curvature inversion is observed briefly just after deposition, but quickly gives way to a steady state with no inversion. We show the essential role of vaporization in shaping the underbelly of the hydrogel, where  changes due to direct mass loss are more significant than the balance of vapor pressure and elastic forces.   

\end{abstract}

\maketitle


A liquid droplet deposited on a hot surface just above its boiling point evaporates quickly and violently. However, at much higher temperatures, an unexpected phenomenon occurs: the droplet levitates and remains floating above the surface for extended periods. This counterintuitive behavior is known as the Leidenfrost effect, and it arises due to the formation of a vapor layer beneath the droplet. This vapor layer insulates the droplet from the surface and simultaneously supports its weight, allowing it to levitate~\cite{biance_leidenfrost_2003, quere_leidenfrost_2013, dupeux_trapping_2011}. The Leidenfrost effect provides a fundamental framework for understanding fluid–structure interactions, with industrial applications ranging from controllable wetting~\cite{wakata_how_2023, jiang_inhibiting_2022, shirota_dynamic_2016} and optimized heat exchange~\cite{wells_sublimation_2015, tran_drop_2012, grounds_enhanced_2012} to high-temperature spray cooling in metallurgical processes~\cite{ravikumar_EnhancementSprayCooling_2023,liang_ReviewSprayCooling_2017}. Leidenfrost physics has been extensively studied in simple liquids~\cite{celestini_take_2012, bouillant_leidenfrost_2018,limbeek_leidenfrost_2017,graeber_leidenfrost_2021,chantelot_leidenfrost_2021, shi_leidenfrost_2025, linke_self-propelled_2006, ma_star-shaped_2017} and sublimable solids~\cite{cousins_ratchet_2012, lagubeau_leidenfrost_2011, dupeux_self-propelling_2013, dorbolo_spontaneous_2016}, where both experimental~\cite{  burton_geometry_2012, bouillant_symmetry_2018, vakarelski_stabilization_2012, hidalgo-caballero_leidenfrost_2016, pacheco-vazquez_triple_2021, bouillant_thermophobic_2021, pacheco-vazquez_designing_2024, paulovics_leidenfrost_2024} and theoretical~\cite{snoeijer_maximum_2009, baier_propulsion_2013, sobac_leidenfrost_2014, sobac_small_2025} efforts have characterized droplet morphology and surface temperature profiles in relation to the droplet’s ability to float. More recently, this phenomenon has been extended to vaporizable soft solids, such as hydrogel spheres, which exhibit two distinct types of Leidenfrost behavior~\cite{waitukaitis_coupling_2017, waitukaitis_bouncing_2018}. At high approach velocities, these spheres make contact with the surface, leading to high-frequency surface oscillations that harvest energy and lead to sustained bouncing~\cite{waitukaitis_coupling_2017}. At low velocities, they enter a ‘normal’ floating state that, in many respects, resembles the classical liquid Leidenfrost state~\cite{waitukaitis_bouncing_2018}. 

The morphology beneath Leidenfrost droplets arises from a balance between surface tension, vapor pressure, and gravity. Burton \textit{et al.}~investigated this balance experimentally using high-speed interferometric imaging through a transparent, heated substrate~\cite{burton_geometry_2012}. Their results revealed a rim with saddle points and extrema, indicating that the lower facet of the droplet is not flat but exhibits a curvature inversion---\textit{i.e.}, a ‘pocket’ of vapor exists below the droplet. This shape had been predicted by Snoeijer \textit{et al.}, whose analytical work showed that the curvature inversion becomes more pronounced with increasing droplet radius~\cite{snoeijer_maximum_2009}. For soft vaporizable solids, one might imagine that the shape of the underbelly should reflect a balance between elastic forces and vapor pressure. However, unlike liquid droplets, where material can always flow to acheive energetic balance, solids can only deform (elastically or plastically) and lose material---they have no flow internally. Hence fundamental differences between the observed shapes in the two cases can be expected.

In this work, we experimentally measure the geometry underneath a Leidenfrost-floating soft solid. We find that the curvature inversion predicted for elastic objects~\cite{binysh_modeling_2023} is lost over time due to vaporization-induced shape changes. An initial elastic inversion forms as the hydrogel sphere approaches the hot surface, but rapidly transitions to a steady state without a vapor pocket and with a mostly flat interface. We propose that this transition results from irreversible deformation of the sphere caused by continued vaporization. Using simulations, we qualitatively reproduce our observations. Our results emphasize the central role of vaporization in altering the shape of Leidenfrost-levitated solids, a mechanism that is fundamentally different from that in liquids.

\begin{figure}[ht]
\includegraphics[scale=1]{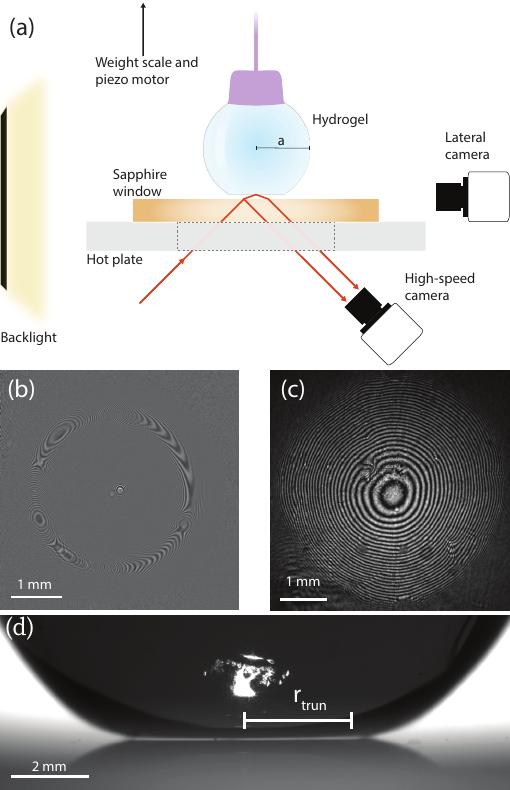}
\caption{\label{fig:Setup}~Experimental setup and key observations. (a)~We use a piezo motor to slowly lower a hydrogel sphere connected to a string toward a hot ($\sim$220$^\circ$C) sapphire window. A weight sensor at the top of the string allows us to determine how much of the hydrogel is supported by the vapor layer. The hot plate below the sapphire window has an aperture that allows an expanded, 633 nm laser to pass through. A high-speed camera records the interference pattern produced from reflections at the top of the window and the bottom of the hydrogel, while a second camera records from the side. (b)~When a water droplet is placed on the window, we observe a steady state with a rim pattern and both saddle/minima features, indicating a curvature inversion beneath the droplet. (c)~In contrast, the steady-state pattern for a hydrogel only has concentric rings, indicating the absence of curvature inversion. (d)~View of the gap below the hydrogel as seen by the side-view camera, with truncated radius  ($r_{\text{trun}}$) indicated.}
\end{figure}



The experimental setup is shown in Fig.~\ref{fig:Setup}(a).  We use a suction cup to attach a hydrogel sphere~(radius $a\approx 7$ mm) to a string, which is connected to a weight sensor and piezo motor. The hot surface below the hydrogel is a wedged sapphire window, chosen for its good heat conduction and transparency. We lower the hydrogel toward the substrate at a constant speed~($\sim$33 \textmu{m}/s) with the piezo motor~(Agilis AG-LS25). The window sits on a hot plate, with an aperture that permits an expanded, 633 nm laser~(Thorlabs HNL210L) to pass through. This beam reflects off the upper surface of the window and the lower surface of the hydrogel to create an interference pattern, which is recorded by a camera~(Phantom v1612). A second camera~(Basler a2A 1920) simultaneously records from the side.
\begin{figure}[ht]
\includegraphics[scale=1]{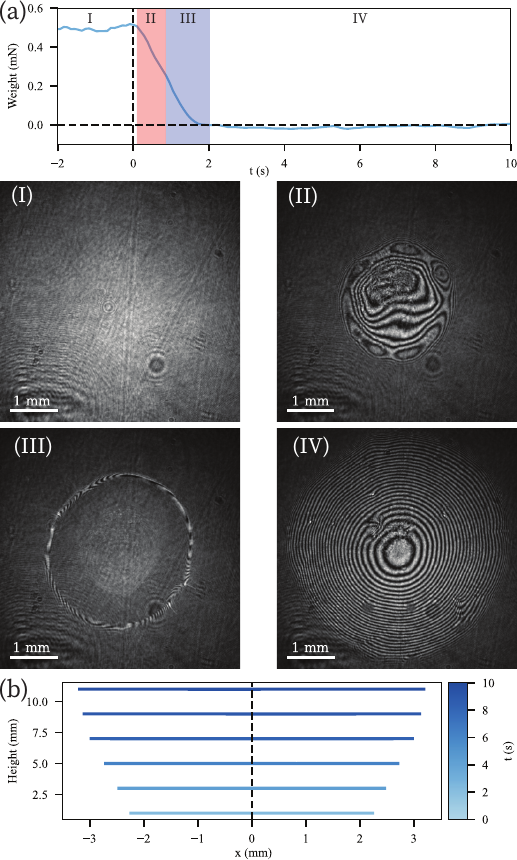}
\caption{\label{fig:Results}~Time evolution of the interference pattern. (a)~Weight of the hydrogel supported by the vapor as a function of time, with key points indicated (I-IV) and corresponding interferometric images. In region I, the weight is carried fully by the string, and no features are present in the interference pattern. At $t=0$, the hydrogel starts interacting with the vapor (region II), and the interference pattern shows a distinct rim with saddle/minima points, indicating curvature inversion. In region III, the rim destabilizes and the underside of the hydrogel enters into oscillations. On region IV, the oscillations disappear, resulting in stable floating of the hydrogel and a stable interference pattern without any curvature inversion. (b)~Reconstructed height profile of the hydrogel underside in region IV at different times, indicating it is essentially flat over.  As we show in the Supplemental Material, there is in fact a very slight upward curvature \cite{Supplemtental_Info}.}
\end{figure}

Figure~\ref{fig:Setup}(b) first shows the steady-state interference pattern of a liquid droplet. A prominent rim of saddle and extrema points indicates the well-known curvature inversion~\cite{burton_geometry_2012}. In the case of a soft solid, one naturally suspects that the interplay of elasticity and vapor pressure should create a similar inversion, as our recent model has indicated~\cite{binysh_modeling_2023}. Our experiments, however, reveal that such reasoning prevails only briefly. As shown in Fig.~\ref{fig:Setup}(c), the steady state for a floating hydrogel has no rim of saddles/minima, but rather just interference pattern of concentric rings, which indicate the absence of curvature inversion. By comparing the radius at which the rings are no longer visible to the `floating' radius, $r_{\text{trun}}$, of the hydrogel seen from the side-view camera (Fig.~\ref{fig:Setup}(d)), we conclude that the absence of an inversion rim is not due to any limit in the field of view of our interferometric images (see Supplemental Material~\cite{Supplemtental_Info}).  

In Fig.~\ref{fig:Results}, we present the time evolution towards this steady state in detail. Panel (a) shows the weight of the hydrogel, supported by the vapor layer as a function of time, with key regimes indicated by Roman numerals (I-IV) and colors. In regime I, the weight is fully supported by the string, indicating that no vapor layer has formed; hence no interference features are present (Fig.~\ref{fig:Results}.(a) image I). At $t=0$, the hydrogel just begins interacting with the vapor, and in the red region (II) we observe a curvature-inversion rim (image II). We only observe this inversion very briefly---in the subsequent blue region (III) the rim destabilizes, giving way to rapid (kHz) oscillations that are accompanied by audible whistling. These oscillations are also short lived; after the weight is fully supported by the vapor, the stable regime (IV) is reached, where only concentric rings are present and there is no inversion. In this regime, the floating radius grows steadily with time, and the number of rings steadily increases (Supplemental Video 1~\cite{Supplemtental_Info}). Fig.~\ref{fig:Results}(b) shows the reconstructed height profiles in the region IV at different times, as determined by extracting the radial extent of each ring and assigning a height difference of one half wavelength. Due to the fact that absent inversion downward curvature is not possible, we conclude that the stable curvature is upward but nearly flat (Fig.~\ref{fig:Results}(b), however with height variations on the micrometer scale over millimeter scale that defines the `floating' radius (see Supplemental Materials \cite{Supplemtental_Info}).

What gives rise to the observed behaviors? When the hydrogel first approaches the surface, its undeformed shape is  spherical, and its initial interaction with the vapor leads to the brief appearance of the anticipated elastic curvature inversion~\cite{binysh_modeling_2023}. This is, apparently, unstable, resulting in the oscillations observed in regime III of Fig.~\ref{fig:Results}. We can explain the subsequent flattening through interactions with the hot surface, which cause not only reversible elastic shape changes but also irreversible changes due to vaporization. As soon as the sphere starts vaporizing, it is no longer a sphere, as every parcel of vapor lost permanently alters the shape. Moreover, as elastic forces and vapor pressure conspire to create the initial curvature inversion and oscillations, the edge of the floating region is on average closer to the surface than the center, hence it vaporizes faster and disappears.
Once the profile is rather flat, it seems that a second mechanism must kick in that maintains a higher vaporization rate at the edges than the center to cause the very slight upward curvature.  We suspect this is due to the hydrogel cooling the substrate locally, though other factors may contribute \cite{Supplemtental_Info}. The takeaway is that elastic deformations must also occur throughout, the dominant contribution to the final shape arises from irreversible mass loss due to vaporization. 
\begin{figure}[ht]
\includegraphics[scale=1]{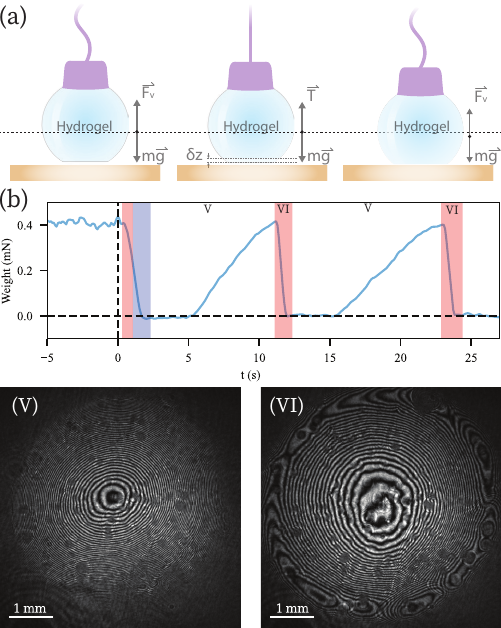}
\caption{\label{fig:Reloading}~Temporary recovery of curvature inversion via elastic reloading. (a)~To re-load the hydrogel elastically, and hence temporarily recover curvature inversion, we start by lowering it just beyond the point where all weight is carried by the vapor, leaving the string slightly loose (left). We then let vaporization occur until a small amount of hydrogel, $\delta z$, is removed, resulting in the string again becoming taut (middle). Upon lowering the hydrogel again, the vapor elastically deforms the hydrogel to recover curvature inversion, but only temporarily as vaporization quickly acts to take it away (right). (b)~Weight \textit{vs.}~time during multiple iterations of the elastic reloading process. Region V corresponds to the recovery of the weight by the string (middle panel a). Region VI, within the red box, highlights the elastic interplay between the hydrogel and the vapor layer. Once again, evaporation overcomes the curvature inversion and the hydrogel reaches a stable floating regime without curvature inversion.}
\end{figure}


We lend evidence to these ideas with the experiment sketched in Fig.~\ref{fig:Reloading}(a). We take a new hydrogel and begin by lowering it toward the hot surface. As soon as its weight is fully carried by the vapor, we stop the piezo motor, which leaves a slight slack in the string. Now we let the sphere float, causing a small vertical slab of thickness $\delta z$ to be removed due to vaporization. When this exceeds the slack, the string starts carrying the weight, the gap becomes larger, and the vaporization rate is reduced. This leads to an elastic unloading of the sphere and a decoupling from the vaporization-dominant mechanism. When the weight has been fully recovered by the string, we lower the sphere again to re-load it elastically and re-initiate the vaporization mechanism, with the prediction that we should (fleetingly) recover the curvature inversion. As shown in Fig.~\ref{fig:Reloading}(b), this prediction bears true. The upper plot shows the weight carried by the vapor throughout the process, where the transference between the vapor/string for multiple iterations is clearly visible. The images below show the repeatable recovery of the curvature inversion.  In V, \textit{i.e.}, after the initial inversion is lost and the sphere has been floating, there are only concentric rings.  However, in VI, just after we have elastically reloaded the sphere, the inversion ring reappears---only to quickly be vaporized again.  (For the full evolution, see Supplemental Video 2~\cite{Supplemtental_Info}.)  

\begin{figure}[ht]
\includegraphics[scale=1]{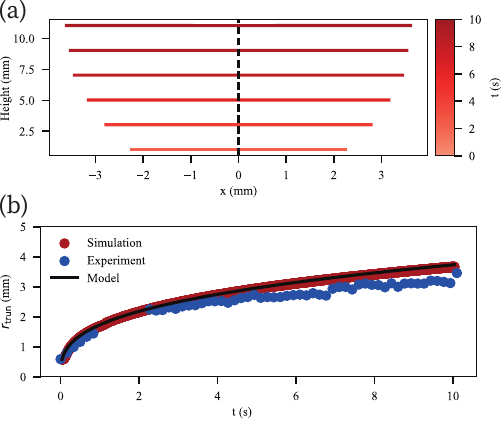}
\caption{\label{fig:Simulations}~Simple vaporization model. 
(a)~Height profile underneath the sphere \textit{vs.}~time, calculated numerically by iteratively applying Eq.~\ref{eq:Evaporation} to determine the evolution of the height/radius profile, Eq.~\ref{eq:PRL_Scott} to determine the corresponding gap height, and FEM simulations to obtain the updated temperature profile.  
(b)~Truncated radius obtained at each iteration of the numerical calculation, compared with the experimentally measured truncated radius and the model prediction. See Supplemental Material~\cite{Supplemtental_Info} for full simulation details.}
\end{figure}

We recover most aspects of the stable regime (IV, Fig.~\ref{fig:Results}(a)) using a simple simulation approach, which involves three steps: (1) using finite element method (FEM) simulations to calculate the temperature profile of the substrate underneath the hydrogel with average gap height $h_0(t)$ and floating radius $r_{trun}(t)$, (2) using this temperature profile to numerically calculate the changes in the height profile (and consequently, floating radius) considering the local temperature gradient and a small time interval, $\delta t$, and (3) use the known force-balance equation to update the average gap height, $h_0$, based on the floating radius. In all steps, we take advantage of the fact that the deviations in the height profile (\textit{i.e.}, the slight upward curvature) are much smaller than the average gap height. Details on step 1, \textit{i.e.}~the FEM simulations, can be found in the Supplemental Material~\cite{Supplemtental_Info}. Step two is achieved using the mass-loss equation,
\begin{equation}
\delta h(r) = - \frac{\kappa}{\rho L} \frac{T(r)-T_H}{h(r)}\delta t,
\label{eq:Evaporation}
\end{equation}
where $\kappa$ is the thermal conductivity of the vapor, $\rho$ and $L$ are, respectively, the density and latent heat of the hydrogel, $T(r)$ is the temperature at radius $r$ on the substrate below, and $T_H=100^\circ$ C is the temperature of the hydrogel.  Step three is achieved using the force balance equation for the integrated lubrication pressure \textit{vis-\`a-vis} the weight of the sphere, given by
\begin{equation}
h_0(t) = \left(\frac{ 3\pi \eta \kappa \Delta T}{2 \rho L} \right)^{\frac{1}{4}} \frac{r_{trun}(t)}{(mg)^{1/4}}.
\label{eq:PRL_Scott}
\end{equation}
Here, $\eta$ is the viscosity of the vapor, $\Delta T$ is the (average) temperature difference between the hydrogel and substrate, and $mg$ is the weight of the hydrogel. Though this equation was originally derived for the case of liquid droplets, it was recently shown to be in good agreement for floating hydrogels~\cite{waitukaitis_bouncing_2018}. 


Results are shown in Fig.~\ref{fig:Simulations} and the corresponding Supplemental Video~\cite{Supplemtental_Info}. Panel (a) shows the (relative) height profiles at different times. Similar to the experiments, the model produces an underbelly hydrogel profile that is \textit{almost} perfectly flat, but with a slight upward curvature (which is comparable to the experimental observations; see \cite{Supplemtental_Info}). In each iteration of the numerical calculation, the truncated radius is extracted from the resulting height profile; this truncated radius then defines the input conditions for the next iteration. Thus, the truncated radius can be used as a control parameter for the simulations. Fig.~\ref{fig:Simulations}(b) shows that the simulated truncated radius (red points) is in agreement with both the experiments (blue points) and the model prediction (solid black line)~\cite{waitukaitis_bouncing_2018}.

Our results underscore the dominant role of mass loss due to vaporization in shaping the long-term height profile beneath a Leidenfrost-floating elastic solid. In hindsight, this makes perfect sense---liquids can flow, and therefore they always will flow until they reach equilibrium. In the case of liquid Leidenfrost droplets, this is set by balancing pressure in the vapor layer, surface tension, and gravity. Even though a liquid droplet loses mass during vaporization, flow is what allows it maintain its equilibrium shape.  Solids, on the other hand, cannot do this.  Although curvature inversion emerges in the early stages of a hydrogel sphere’s interaction with a hot surface---and can be briefly re-initiated via elastic reloading---it rapidly disappears as irreversible, vaporization-induced deformations proceed. The solid cannot flow to maintain a minimal energetic state. 
The simulations show that vaporization is the main mechanism of mass loss and successfully reproduce the key qualitative features of the observed height profile evolution. While our study emphasizes vaporization-driven shape changes, the remnant influence of elasticity remains incompletely explored. A complete analytical treatment of the system---including squeeze flow in the vapor layer, elastic deformations, vaporization dynamics, substrate cooling, and the resulting force balances---is highly complex and beyond what we can assess here. These coupled processes may be more effectively studied using multi-physics simulations to capture all aspects. 

This research was supported by the Scientific Service Units of The Institute of Science and Technology Austria (ISTA) through resources provided by the Miba Machine Shop, and the Scientific Computing Facility.
J.B.~acknowledges funding from the European Union’s Horizon research and innovation programme under the Marie Sklodowska-Curie Grant Agreement No.~101106500.

V.D.~and I.L.~contributed to the development of the methodology, formal analysis, and software development. J.B.~and A.S.~contributed to the conceptualization. V.D.~and S.W.~contributed to writing and editing the manuscript. S.W.~supervised the research activity. 

\bibliographystyle{apsrev4-2}
\bibliography{PRL_Hydrogel_2025}

\end{document}


\maketitle

\section{Upward curvature underbelly hydrogel}
\begin{figure}[ht]
\centering
\includegraphics[scale=1]{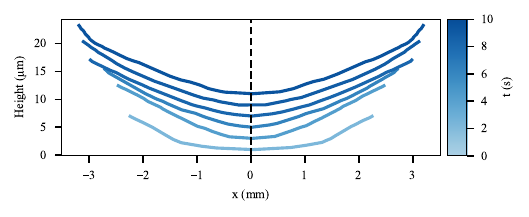}
\caption{\label{fig:Upward_exp}Zoomed-in view of the experimental profiles extracted from the interference patterns at multiple times. }
\end{figure}

As mentioned in the main manuscript, when the hydrogel transitions from the inverted curvature at the interface between the hydrogel and the vapor layer to a mostly flat surface, the tendency is toward a slight upward curvature. Fig.~S\ref{fig:Upward_exp} shows a zoomed-in view of the experimental profiles presented in Fig.~2(b) of the main manuscript.

\section{Finite elements simulations}

\begin{figure}[ht]
\centering
\includegraphics[scale=1]{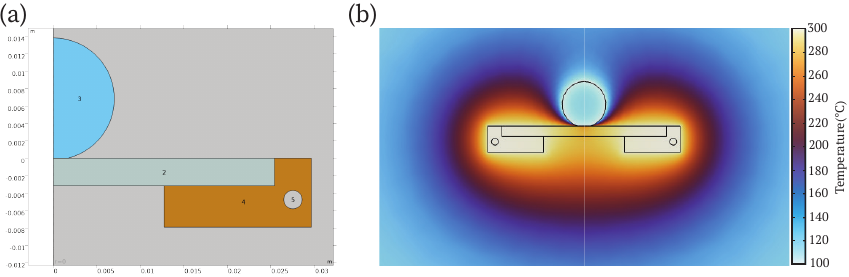}
\caption{\label{fig:comsol_geometry} (a)~Geometry used in the finite element method (FEM) simulations. We performed 2D axisymmetric simulations using \textsc{COMSOL Multiphysics}, in which a truncated hydrogel sphere is placed above a sapphire substrate. (b)~Results of the FEM simulations showing the system temperature for different truncation radii of the sphere. }
\end{figure}

We use the finite element method (FEM), implemented in \textsc{COMSOL Multiphysics} (version 6.1), to calculate the radially dependent temperature profile of the substrate beneath the hydrogel. We perform a 2D axisymmetric simulation by placing a truncated hydrogel sphere above a sapphire surface, as shown in Fig.~S\ref{fig:comsol_geometry}(a).

The gap height corresponds to the truncation radius and follows the relation:
\begin{equation}
h_0(t) = \left(\frac{3\pi \eta \kappa \Delta T}{2 \rho L} \right)^{\frac{1}{4}} \frac{R(t)}{(mg)^{1/4}}.
\label{eq:PRL_Scott}
\end{equation}
This relation is valid for both liquid droplets and hydrogels. The sapphire substrate is in contact with a copper heater. We solve the heat transfer equation for the geometry described above, applying the following boundary conditions: internal boundaries in contact with the heater are set to $300^\circ\mathrm{C}$, while the boundaries of the hydrogel are maintained at $100^\circ\mathrm{C}$. The surroundings of both the hydrogel and the heater are modeled as air, with thermally insulating (adiabatic) boundary conditions.

The initial conditions assume the hydrogel is at $100^\circ\mathrm{C}$, and both the sapphire and heater are at $300^\circ\mathrm{C}$. A physics-controlled mesh with an extremely fine element size is used to ensure numerical accuracy.

\begin{figure}[ht]
\centering
\includegraphics[scale=0.9]{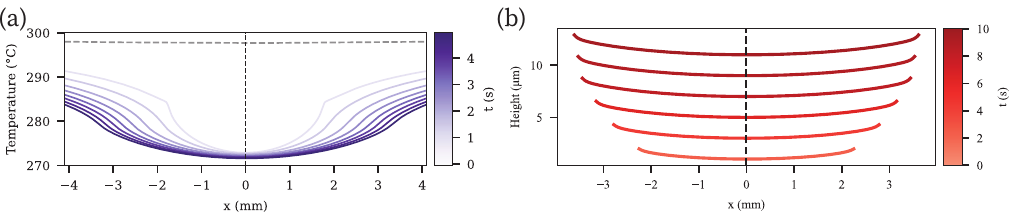}
\caption{\label{fig:Temp_profile} (a)~Steady-state temperature profiles at the top surface of the sapphire window for varying truncated radii. (b)~Zoomed-in view of the underbelly height profile of the hydrogel, numerically calculated using the temperature profiles extracted from the FEM simulations.}
\end{figure}

We perform a stationary study for different truncation radii, yielding the temperature profiles used in the local mass loss calculations described in the main text (Fig.S\ref{fig:Temp_profile}~(a)).

Fig.~S\ref{fig:Temp_profile}(b) shows the resulting zoomed-in height profile obtained from the numerical calculations of our system. We observe that the results are qualitatively similar to the experiments, exhibiting an upward curvature of the same order of magnitude.



\section{Field of view of our interferometry setup}
\begin{figure}[t]
\centering
\includegraphics[scale=1.5]{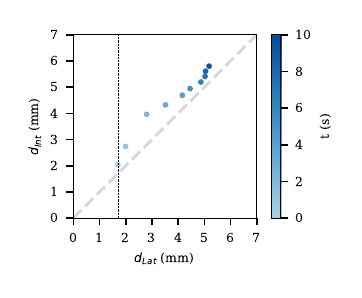}
\caption{\label{fig:diameter_diameter} Lateral diameter measured from the side-view camera image versus the diameter obtained from the interference pattern. The strong agreement between the two methods indicates reliable access to the optical features beneath the hydrogel}
\end{figure}

We measured the lateral diameter of the truncated hydrogel in two different ways: (1) from the side using a camera, and (2) by analyzing the diameter of the interference pattern. The side-view image of the hydrogel was fitted using a piecewise function, where the curved region was approximated with a parabola and the truncated section with a straight line. The distance between the intersections of the two parabolic fits was taken as the lateral diameter in the side view. In the case of the interference pattern, the diameter was measured directly from the pattern. Fig.S\ref{fig:diameter_diameter} shows that the two measurements closely match, providing evidence that we can reliably access all the optical features beneath the hydrogel with our interferometry setup; \textit{i.e.}, the absence of the inversion ring is not due to a limited field of view.

\section{Height variation calculation}

The mass loss process is a key component that defines the final shape of the hydrogel underbelly. In our system, vaporization-induced shape changes can originate mainly from two sources: the spatially dependent heat flux and the concentration gradients in the vapor layer.

\subsection{Heat flux contribution}
For the heat flux ($\vec{J_q}$), heat conduction is dominant. We use Fourier's Law (Equation~\ref{eq:Fourier_law}) to calculate the local variation in height as a consequence of the temperature gradient $\nabla T$, where $\kappa$ is the thermal conductivity:
\begin{equation}
    \vec{J_q} = - \kappa \nabla T.
    \label{eq:Fourier_law}
\end{equation}
Assuming temperature changes only along the $z$-axis for local vaporization and adapting to polar coordinates, we derive Equation~\ref{eq:vaporization_heat}, where $L$ is the latent heat of vaporization:
\begin{equation}
    \delta M = - \frac{\kappa}{L} \frac{\partial T}{\partial z} 2\pi r \, dr \, \delta t,
    \label{eq:vaporization_heat}
\end{equation}
Here, $\delta M$ is the mass variation over the time interval $\delta t$ along the truncated region of the hydrogel. Using the definition of density $\rho$, we relate mass variation $\delta M$ to the local height variation $\delta h$ (Equation~\ref{eq:mass_variation}):
\begin{equation}
    \delta M = \rho \, dV = \rho \, 2\pi r \, dr \, \delta h.
    \label{eq:mass_variation}
\end{equation}
Combining Equations~\ref{eq:vaporization_heat} and~\ref{eq:mass_variation}, and applying the thin-layer approximation, we obtain the height variation as a function of time (Equation~\ref{eq:delta_differential_h}). Here, $h(r)$ is the absolute height at a radial position $r$, $T(r)$ is the radial temperature profile, and $T_H$ is the hydrogel temperature:
\begin{equation}
    \delta h = - \frac{\kappa}{L\rho} \frac{T(r)-T_H}{h(r)} \delta t.
    \label{eq:delta_differential_h}
\end{equation}

\subsection{Vapor concentration flux contribution}
\begin{figure}[h]
    \centering
    \includegraphics[width=0.65\linewidth]{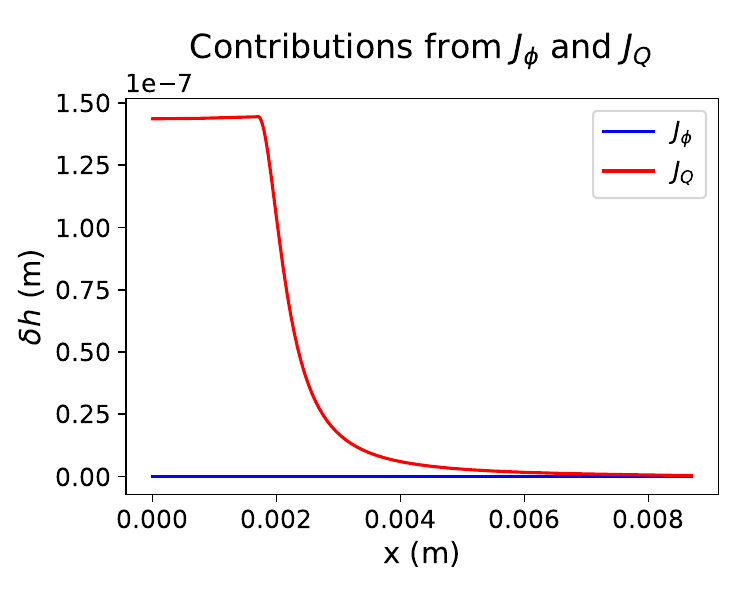}
    \caption{Contribution to the height variation($\delta h$) from the heat flux($J_Q$) and vapour concentration flux ($J_{\phi}$).}
    \label{fig:enter-label}
\end{figure}

A second possible contributor to the slight upward curvature observed in the steady-state is variable vapor flux across the hydrogel bottom.  We can estimate the contribution of this using Fick's Law (Equation~\ref{eq:Fick_law}) to relate the concentration flux ($\vec{J_\phi}$) with the concentration gradient ($\nabla \phi$), where $D$ is the diffusion coefficient:
\begin{equation}
    \vec{J_\phi} = - D \nabla \phi.
    \label{eq:Fick_law}
\end{equation}
By applying the definition of concentration flux in polar coordinates, and using the relation between mass and concentration via molar mass ($M$), together with Equation~\ref{eq:mass_variation}, we derive the height variation due to concentration gradient (Equation~\ref{eq:delta_differential_h_concentration}):
\begin{equation}
    \delta h = - \frac{D M}{\rho} \nabla \phi \, \delta t.
    \label{eq:delta_differential_h_concentration}
\end{equation}
From the ideal gas law, we relate concentration $\phi$ to pressure $P$, where $R$ is the ideal gas constant and $T$ is the temperature:
\begin{equation}
    \phi = \frac{P}{R T}.
    \label{eq:Ideal_gas}
\end{equation}
From lubrication theory, the pressure profile underneath the hydrogel is given by Equation~\ref{eq:Lubrication_pressure}, where $\mu$ is the viscosity, $h_0$ is the absolute current height, $P_{\text{atm}}$ is atmospheric pressure, $r_{trun}$ is the truncated radius, and $r$ is the radial position:
\begin{equation}
    P(r) = \frac{3\mu W_0}{h_0^3}(r^2 - r_{trun}^2) + P_{\text{atm}},
    \label{eq:Lubrication_pressure}
\end{equation}
$W_0$ is defined as:
\begin{equation}
    W_0 = \frac{\kappa (T(r) - T_H)}{\rho L h_0}.
    \label{eq:W0}
\end{equation}
Combining Equations~\ref{eq:delta_differential_h_concentration},~\ref{eq:Ideal_gas},~\ref{eq:Lubrication_pressure}, and~\ref{eq:W0}, we obtain the height variation due to concentration flux as:
\begin{equation}
    \delta h = - \frac{6 D M \mu \kappa}{\rho^2 R T_H L h_0^4} (T(r) - T_H) r \, \delta t.
    \label{eq:delta_height_radius}
\end{equation}
Following the computational steps described in the main text, we calculate the height variation over discrete time steps. Figure~S\ref{fig:enter-label} shows the results for the initial stages, where it is evident that the heat flux has a more significant impact on height variation compared to the concentration flux.

\subsection{Contribution of dried hydrogel `shell'}
A third possible contribution to the small observed upward curvature is elastic restoring forces from the thin `shell' of water-free hydrogel that remains after water has been removed. One can estimate its thickness by a back-of-the-envelope calculation. In the longest experimental run presented in the paper, the amount of water vaporized corresponds to a spherical cap approximately 2 mm in height at the bottom of the hydrogel. Given that our hydrogels are $>\, 99.9\%$ water by mass, the solid polymer ``skin'' remaining from this cap is approximately $2\,\text{mm} \times (1 - 0.999) \sim 1\,\mu\text{m}$ thick. Additionally, as typical polymers have a Young's modulus on the order of $\sim 1$ GPa,  we expect this shell to have a Young's modulus of the same order of magnitude. Although predicting the exact influence of this shell on the observed shape is beyond our scope, qualitatively it is expected to act as a stiff but stretchable membrane that pulls the evaporated region toward the center. Hence, it is possible that this effect also contributes to the slight upward curvature.

%
